\documentclass[%
 aip,
 apl,
 amsmath,amssymb,
 reprint,%
]{revtex4-1}

\usepackage[utf8]{inputenc}
\usepackage{color}
\usepackage{soul}
\usepackage[alsoload=synchem]{siunitx}
\usepackage[version=3]{mhchem}
\usepackage{graphicx}
\usepackage{dcolumn}
\usepackage{bm}
\usepackage[acronym,shortcuts]{glossaries}
\usepackage[percent]{overpic}
\usepackage{url}
\usepackage[breaklinks,colorlinks=true,allcolors=blue]{hyperref}
\usepackage{subfigure}

\definecolor{purple}{rgb}{0.39, 0.0, 0.39}

\makeatletter
\def\@email#1#2{%
 \endgroup
 \patchcmd{\titleblock@produce}
  {\frontmatter@RRAPformat}
  {\frontmatter@RRAPformat{\produce@RRAP{*#1\href{mailto:#2}{#2}}}\frontmatter@RRAPformat}
  {}{}
}%
\makeatother

\begin{document}
\preprint{AIP/123-QED}

\title{Light-shift mitigation in a microcell-based atomic clock with\\ Symmetric Auto-Balanced Ramsey spectroscopy}

\author{M. Abdel Hafiz}
\affiliation{FEMTO-ST, CNRS, Universit\'e Bourgogne Franche-Comt\'e, Besançon, France}
\author{C. Carl\'e}
\affiliation{FEMTO-ST, CNRS, Universit\'e Bourgogne Franche-Comt\'e, Besançon, France}
\author{N. Passilly}
\affiliation{FEMTO-ST, CNRS, Universit\'e Bourgogne Franche-Comt\'e, Besançon, France}
\author{J. M. Danet}
\affiliation{Syrlinks, 28 rue Robert Keller, Cesson-Sevigne, France}
\author{C. E. Calosso}
\affiliation{INRIM, Strada delle Cacce 91, Torino, Italy}
\author{R. Boudot}
\affiliation{FEMTO-ST, CNRS, Universit\'e Bourgogne Franche-Comt\'e, Besançon, France}
\email{moustafa.abdel@femto-st.fr}


\begin{abstract}
The mid-term fractional frequency stability of miniaturized atomic clocks can be limited by light-shift effects. In this Letter, we demonstrate the implementation of a symmetric Auto-Balanced Ramsey (SABR) interrogation sequence in a microcell-based atomic clock based on coherent population trapping (CPT). Using this advanced protocol, the sensitivity of the clock frequency to laser power, microwave power and laser frequency variations is reduced, at least by one order of magnitude, in comparison with continuous-wave (CW) or Ramsey interrogation schemes. 
Light-shift mitigation obtained with the SABR sequence benefits greatly to the clock Allan deviation for integration times between 10$^2$ and 10$^5$ s. These results demonstrate that such interrogation techniques are of interest to enhance timekeeping performances of chip-scale atomic clocks. 

\end{abstract}

\maketitle
The development of low-SWaP (size-weight-power) frequency references with enhanced frequency stability is of crucial importance in a wide range of applications including timing, navigation, positioning, security, communication or scientific systems \cite{Kitching:APR:2018}. In these domains, microwave chip-scale atomic clocks (CSACs) \cite{Lutwak:PTTI:2007, Zhang:JOSAB:2016, Zhang:Ulpac, Vicarini:UFFC:2019, Yanagimachi:APL:2020} based on coherent population trapping (CPT) have met a remarkable success by offering a daily drift about 100 times smaller than commonly-used oven-controlled quartz oscillators.\\
Light-shifts are known to be an important contribution to the fractional frequency stability of miniaturized atomic clocks for integration times higher than 100~s. Multiple approaches have then been proposed to mitigate their detrimental impact. Some efforts were oriented towards the extraction of the actual laser or cell temperature, from the atomic response itself, in order to reduce the negative impact of temperature gradients between these key components and their respective temperature sensors \cite{Gerginov:OL:2006, Lutwak:PTTI:2007, Vicarini:UFFC:2019}. Other sophisticated methods include the active stabilization of a specific laser microwave modulation index that reduces laser power-induced frequency instabilities \cite{ZhuPatent:2001, Shah:APL:2006, McGuyer:APL:2009, Zhang:JOSAB:2016, Vicarini:UFFC:2019}, possibly combined with the compensation for the laser aging \cite{Yanagimachi:APL:2020}, or the implementation of advanced tailored interrogation sequences using laser power modulation \cite{MAH:PRAp:2020}. Deposition of gold micro-discs, used as privileged alkali condensation spots onto the cell windows was also shown to avoid the progressive obstruction of the transmitted laser light \cite{Karlen:EFTF:2018}.\\
An alternative and straightforward approach to mitigate light-shifts in CPT clocks is to probe the clock transition with Ramsey spectroscopy \cite{Zanon:PRL:2005, Castagna:UFFC:2009}. Mainly investigated in compact vapor cell clocks \cite{MAH:JAP:2017}, Ramsey-CPT spectroscopy  has recently stimulated some research interest in miniaturized atomic clocks. The generation of robust Ramsey-CPT sequences with directly-modulated lasers was demonstrated in Refs \cite{Yano:2015, Fukuoka:2019}, while the spectroscopy, clock operation and evidence of light-shift mitigation was reported with Ramsey-CPT in buffer-gas filled microfabricated vapor cells \cite{Boudot:JOSAB:2018,Carle:UFFC:2021}.\\
Nevertheless, Ramsey-CPT spectroscopy suffers from a residual sensitivity to light-shifts, experienced by the atoms during the light pulses. Advanced Ramsey-based tailored interrogation protocols, based on two consecutive Ramsey sequences with different dark times, have then been proposed and demonstrated in various types of atomic clocks for enhanced light-shift mitigation \cite{Sanner:PRL:2018, Yudin:PRAp:2018, 
MAH:APL:2018, 
Shuker:PRL:2019, Shuker:APL:2019, Calosso:2019, Basalaev:PRA:2020}. In standard vapor cell clocks, it was demonstrated that the symmetry of the interrogation was of crucial importance to tackle an atomic memory effect, mainly linked to the limited duty cycle of the probing sequence, and then to optimize light-shift mitigation \cite{MAH:APL:2018, MAH:PRAp:2020}. While these techniques have shown outstanding efficiency, they have not yet been explored in microfabricated cells, as those employed in CSACs.\\
In microcells, the shorter timescales of the light pulse sequence, imposed by the reduced CPT coherence lifetime, raises the question of a possible amplification of the atomic memory effect. In addition, the loss of resolution in the light-shift measurement, as the difference between the applied dark times is decreased, deserves specific attention to ensure that enough precision is kept to provide improved clock frequency stability.\\
\begin{figure*}[t]
\centering
\includegraphics[width=0.99\linewidth]{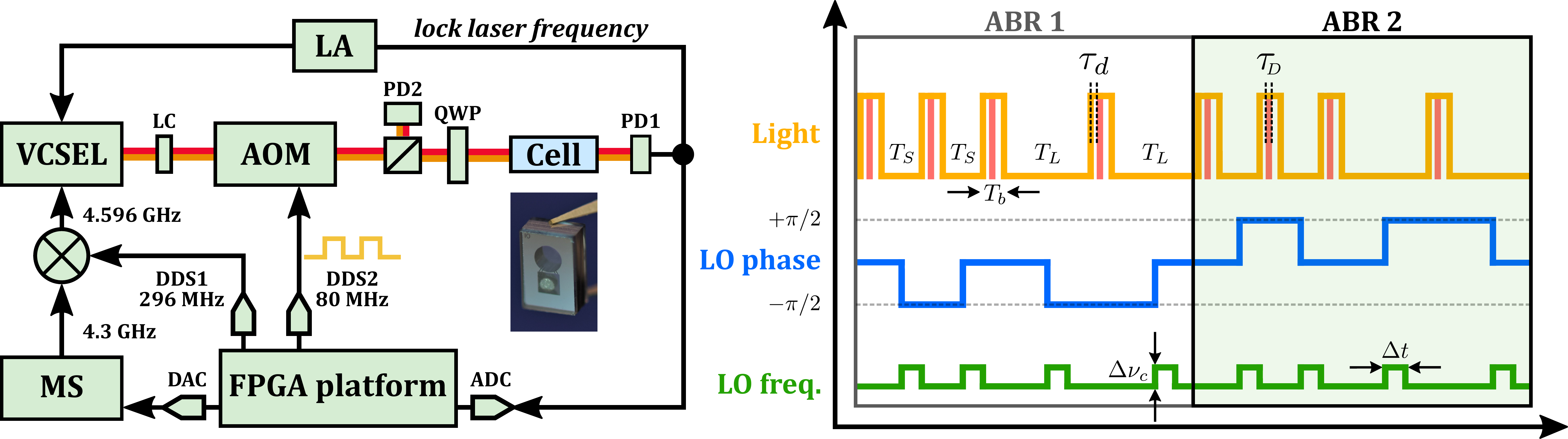}
\caption{(Color online) Left: CPT clock setup and basic architecture of a SABR sequence. LC: collimation lens, QWP: quarter-wave plate, LA: lock-in amplifier, ADC: analog-to-digital converter, DAC: digital-to-analog converter, MS: microwave synthesizer. Right: simplified scheme for the SABR sequence.
}
\label{fig:fig1}
\end{figure*}
In this letter, we study and demonstrate the implementation of a symmetric Auto-Balanced Ramsey (SABR) sequence in a CPT-based microcell atomic clock. For sake of simplicity and proof-of-concept demonstration, the pulsed optical sequence is applied with an external acousto-optical modulator (AOM).
In comparison with the standard Rabi or Ramsey-CPT interrogation schemes, we demonstrate that the SABR method reduces the sensitivity of the clock frequency to laser power, microwave power and laser frequency variations, by a factor higher than 10. Furthermore, we strengthen the importance of the sequence symmetry and its ability to annihilate the consequences of the atomic memory effect in mm-scale cells. We also show that the use of SABR improves the clock Allan deviation for integration times between 10$^2$ and 10$^5$ s, especially by reducing the impact of temperature-induced light-shift effects.\\
Figure \ref{fig:fig1} presents the CPT clock experimental setup. The heart of the clock is a pill-dispenser microfabricated Cs vapor cell \cite{Hasegawa:SA:2011, Vicarini:SA:2018} filled with about 90 Torr of Neon. The cell is temperature-stabilized at 70$^{\circ}$C. A static magnetic field of 10~$\mu$T is applied to raise the Zeeman degeneracy and isolate the 0-0 clock transition. Atoms in the cell interact with a dual-frequency optical field produced by direct microwave modulation of a vertical-cavity surface emitting laser (VCSEL) \cite{Kroemer:AO:2016}, tuned on the Cs D$_1$ line, and such that both first-order optical sidebands induce the CPT resonance. Except when willingly varied for tests, the microwave power that enters the VCSEL is about $-$2.3 dBm. At the output of the laser, an AOM, driven by a switchable radiofrequency (RF) signal, is used to generate the pulsed optical sequence. Tuning the power of the RF signal permits control of the total laser power incident onto the cell. The 0.5-mm diameter laser beam is then sent through the Cs vapor microcell and detected at its output with a photodiode.\\
For this study, we have used a field-programmable gate array (FPGA)-based digital control electronics platform \cite{Sinara}. The latter allows fast computation and feedback to the experiment, such that the generated sequence is readjusted every clock cycle. In this ecosystem, the output 4.596 GHz signal that drives the VCSEL is obtained by mixing a 4.3 GHz from a synthesizer and a 296 MHz signal generated by a direct digital synthesizer (DDS1). A second DDS (DDS2) delivers the 80 MHz RF signal that drives the AOM. This signal can be turned on and off with the help of embedded RF switches. All DDS are clocked with an ultra-pure 860 MHz signal obtained by frequency division by 5 of the synthesizer 4.3 GHz signal. The absolute phase noise of the output 4.596 GHz was measured to be $-$118 dBc/Hz at an offset frequency $f$~=~1 kHz. This phase noise reduces the Dick effect contribution \cite{Danet:UFFC:2014} to a negligible level for such a microcell-based clock. 
The microwave source is referenced to an active hydrogen maser.
In the following, the optically-carried 9.192 GHz signal of frequency $\nu_{LO}$, used to probe the atomic transition, is named as the local oscillator (LO) signal. In clock operation, the value of $\nu_{LO}$ is changed by digitally changing the output frequency of DDS1. Corrections applied to DDS1 are then recorded and used as frequency data for analysis.
\\
The tested SABR sequence, shown in Fig. \ref{fig:fig1}, is similar to the one described in Ref. \cite{MAH:APL:2018}. It consists of two consecutive ABR sequences (ABR~1 and ABR~2). Each ABR sequence consists in turn of four consecutive Ramsey-CPT sequences with light pulses of length $T_b$. The two first Ramsey-CPT patterns use a short dark time $T_S$ while the two following ones use a long dark time $T_L$. A $\pm \pi/$2 phase jump is applied onto the optically-carried microwave interrogating signal during the dark times by acting on DDS1, in order to successively measure the transmitted signals on respective sides of the central fringe. This yields the extraction of error signals, noted $\varepsilon_S$ (for the short dark time pattern) and $\varepsilon_L$ (for the long dark time pattern). In the second ABR sequence (ABR~2), the light pulse pattern is identical while the LO phase modulation pattern is the mirror symmetric to the one used in the first ABR sequence (ABR~1). In vapor cell experiments, the use of a symmetric ABR sequence is of crucial importance to cancel a memory effect of the atoms and then to improve the efficiency of the light-shift rejection \cite{MAH:APL:2018}. Ultimately, two error signals are calculated, ensuring that information is extracted at all pulses for both correction signals and then preventing the negative impact of aliasing on the short-term stability \cite{MAH:APL:2018}. The error signal $\varepsilon_+ = \varepsilon_S + \varepsilon_L$ is used to correct the LO frequency. The error signal $\varepsilon_- = \varepsilon_S - \varepsilon_L$ is extracted to correct the value of an additional phase jump $\varphi_c$, applied during dark times, that compensates for the light-induced phase shift built up during the previous light pulse. The phase jump $\varphi_c$ is obtained by changing the LO frequency by the amount $\Delta \nu_c$ for a time $\Delta t$.\\
In all the tests reported in this manuscript, the pulses length $T_b$ is set to 183 $\mu$s and the atomic signal is sampled after a delay $\tau_d$ of 33 $\mu$s, to account for the delay induced by the anti-aliasing filter. The latter, a 25.6 kHz low-pass filter, averages the photodiode voltage over about 20 $\mu$s and is responsible of the actual duration $\tau_D$ of the detection window. \\

\begin{figure}[t]
\centering
\includegraphics[width=0.95\linewidth]{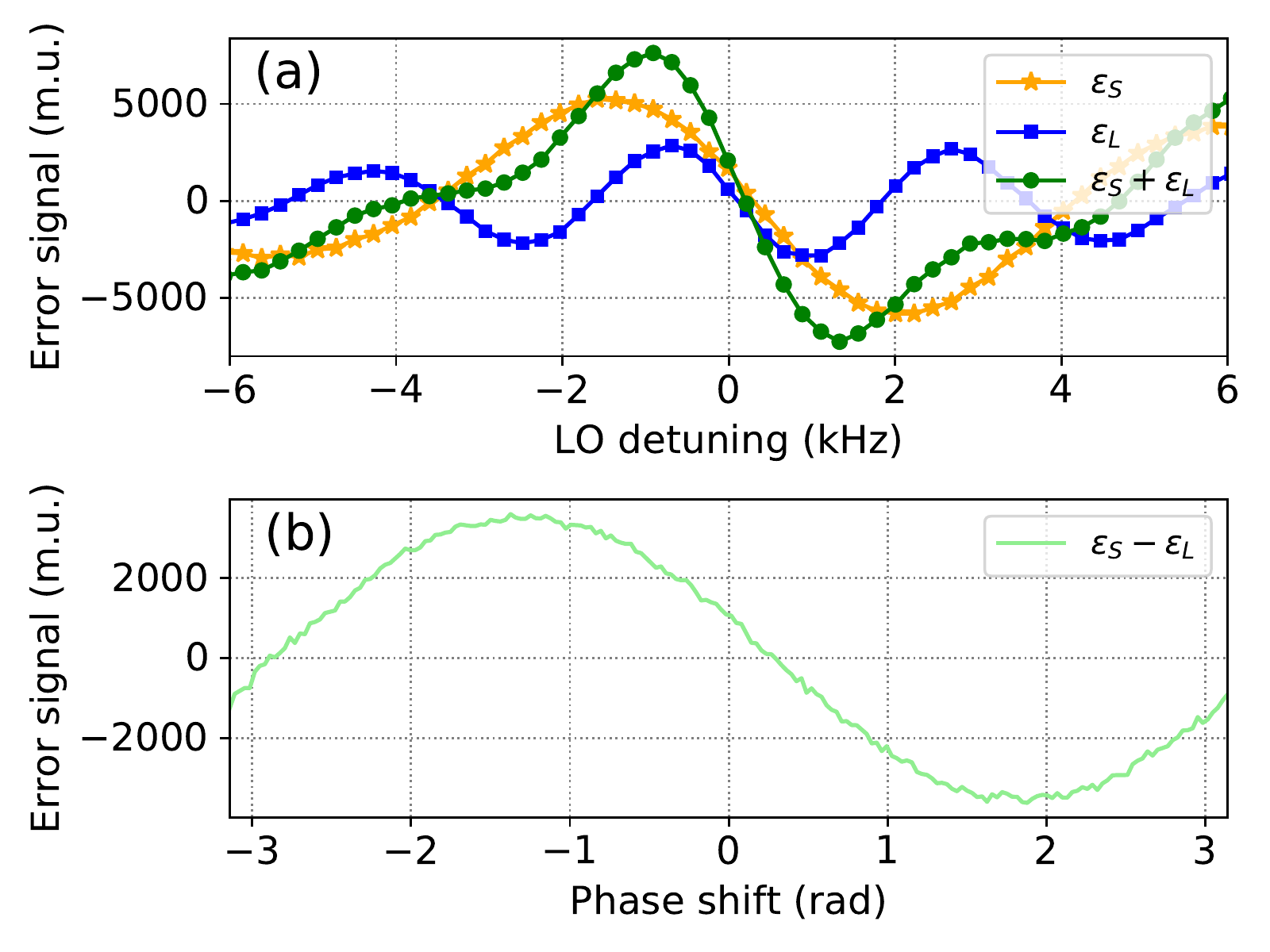}
\caption{(Color online) (a) Error signals $\varepsilon_S$, $\varepsilon_L$, $\varepsilon_+$ as a function of the LO detuning from the value of 9 192 682 200 Hz, and (b) Error signal $\varepsilon_-$ as a function of the phase jump $\varphi_c$ during the dark time, extracted from a SABR sequence. 
}
\label{fig:fig2}
\end{figure}

\begin{figure}[t]
\centering
\includegraphics[width=0.95\linewidth]{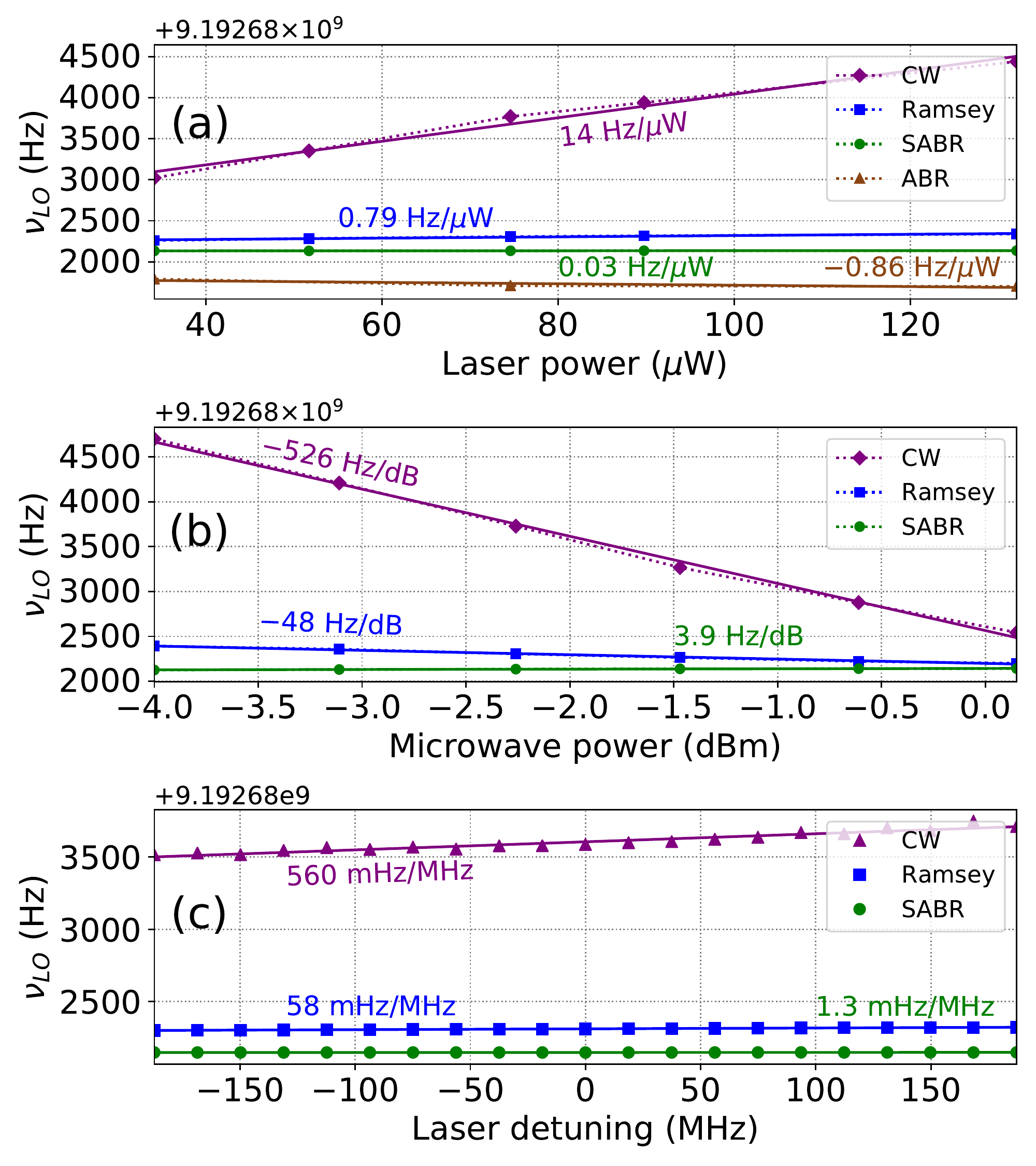}
\caption{(Color online) Clock frequency versus (a) the laser power, (b) the microwave power or (c) the laser carrier frequency, in the standard CW regime, the Ramsey-CPT case, the SABR case or the ABR case. Experimental data are fitted by linear functions, shown as solid lines, from which a linear light-shift coefficient is extracted.
}
\label{fig:fig3}
\end{figure}

\begin{figure}[t]
\centering
\includegraphics[width=0.95\linewidth]{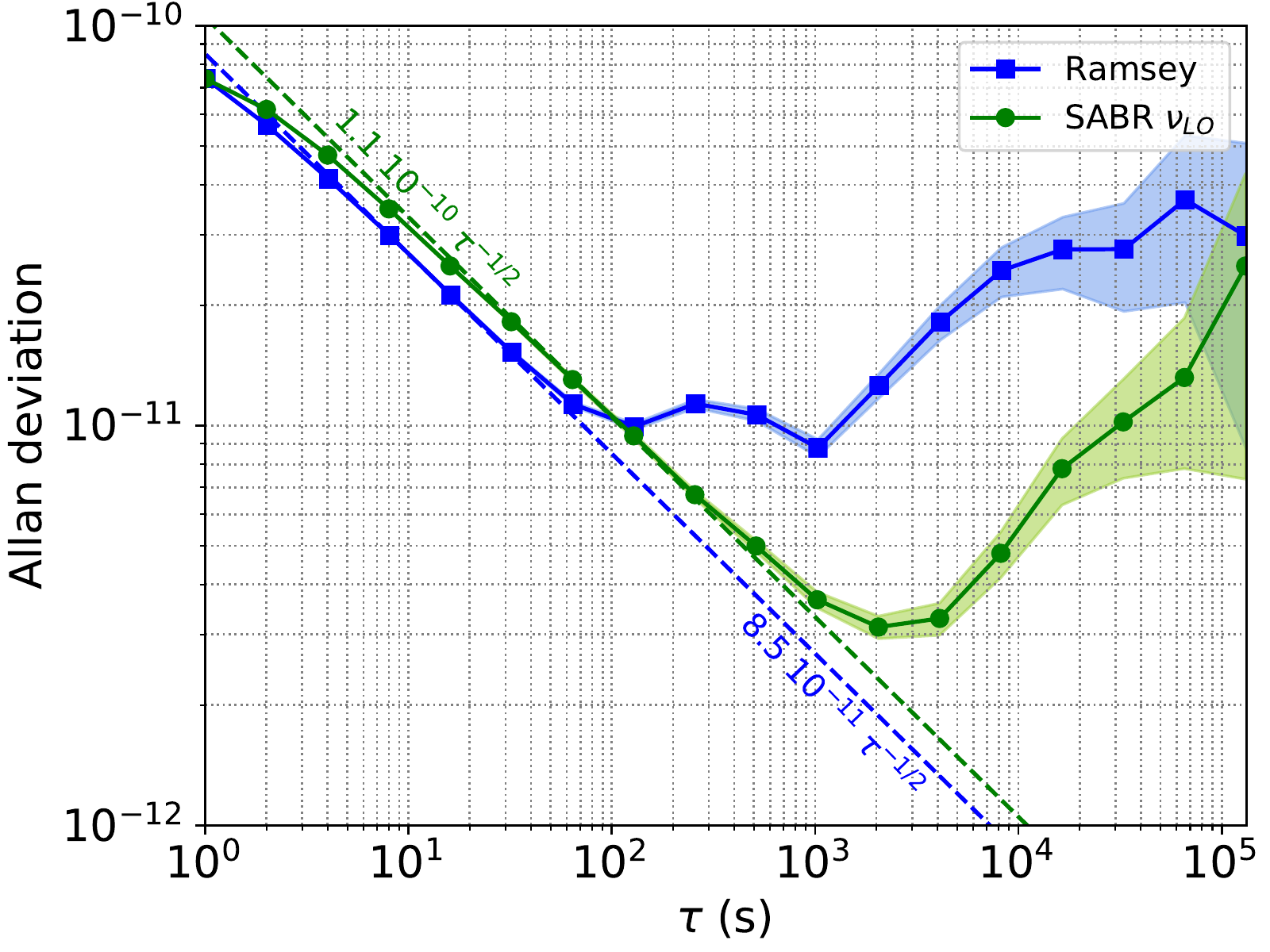}
\caption{(Color online) Allan deviation of the clock frequency in Ramsey-CPT and SABR-CPT regimes.
}
\label{fig:allan_devs}
\end{figure}

Figure \ref{fig:fig2}(a) shows error signals $\varepsilon_{S}$, $\varepsilon_{L}$, and $\varepsilon_{+}$ extracted from a SABR sequence performed on the Cs-Ne microcell. It is clear that the initial error signal $\varepsilon_{S}$, extracted from the sequence with the shortest dark time $T_S$, is higher in amplitude and broader than the error signal $\varepsilon_{L}$ obtained for the longest sequence. Their zero-crossings, mainly shifted from the natural Cs atom frequency because of the buffer-gas induced collisional shift, do not coincide due to the variation of the light-shift magnitude with the dark time value. The error signal $\varepsilon_{+} =  \varepsilon_S + \varepsilon_L$ benefits from a higher amplitude, that justifies its use for stabilization of the LO frequency. The error signal $\varepsilon_{-} =  \varepsilon_S - \varepsilon_L$, plotted in Fig. \ref{fig:fig2}(b) versus the phase jump $\varphi_c$ applied during the dark time, exhibits in open-loop configuration a zero-crossing point at a non-null value of $\varphi_c$ ($\sim$ 0.3 rad), where the light-shift is compensated in closed loop. \\
Figure \ref{fig:fig3}(a) shows the dependence of the clock frequency to laser power variations, in the standard CW regime, the Ramsey-CPT case, the SABR case and the non-symmetric ABR case, respectively. In the Ramsey-CPT case, the light-shift coefficient is 0.79 Hz/$\mu$W. In the SABR case, the latter is reduced by a factor 26 to 0.03 Hz/$\mu$W. This coefficient is 470 times smaller than the one measured in the standard CW scheme (14 Hz/$\mu$W). We have also performed the same measurement when the ABR clock sequence is reduced to the first part (ABR~1). In this non-symmetric case, the memory effect is responsible for a shift of the clock frequency by more than 250 Hz. Without symmetry, the ABR protocol does not give any significant advantage and even leads to a slight deterioration, with respect to the Ramsey-CPT case in microcells. This result attests that the memory effect is strong in microcells, highlighting the importance of the sequence symmetry.\\
Variations of the microwave power that drives the VCSEL can also induce significant shifts of the CPT clock frequency \cite{Zhang:JOSAB:2016, ZhuPatent:2001, Shah:APL:2006}. We have then measured this sensitivity with the SABR sequence. Corresponding results are shown on Fig. \ref{fig:fig3}(b), in comparison with those obtained with the CW or Ramsey-CPT schemes.
In the SABR case, a sensitivity of $+$3.9~Hz/dB is measured, to be compared with a slope of $-$48~Hz/dB and $-$526~Hz/dB in the Ramsey-CPT and CW cases, respectively.\\
Figure \ref{fig:fig3}(c) depicts an additional sensitivity measurement with the laser frequency. Calibration of the laser frequency change was performed using absorption profiles detected at the cell output. Here again, we observe a strong reduction of the light-shift coefficient, reduced from 58 mHz/MHz in the Ramsey-CPT case and even 560 mHz/MHz in CW regime down to 1.3 mHz/MHz in the SABR case.\\ 
Finally, frequency stability measurements of the microcell CPT clock operating with Ramsey-CPT or SABR-CPT sequences have been performed. These tests were conducted on the same setup, with quasi-identical environment conditions. The total laser power at the cell input is about 72 $\mu$W. Allan deviations are shown in Fig. \ref{fig:allan_devs}.\\
In the Ramsey-CPT case, the clock short-term stability is about 8.5 $\times$ 10$^{-11}$ $\tau^{-1/2}$ until 60 s. We identified that the "bump" between 60 and 10$^3$~s was attributed to temperature fluctuations of the laboratory. For $\tau >$~10$^3$~s, a degradation is observed, yielding the level of 3~$\times$ 10$^{-11}$ at 10$^5$ s.\\
In the SABR case, the clock short-term stability is slightly degraded, with the level of 1.1 $\times$ 10$^{-10}$ at 1~s. At the opposite, the clock performance is clearly improved in the 10$^2$ - 10$^5$ s range. We observed with SABR a significant suppression of the correlation of the clock frequency with temperature variations of the laboratory, benefiting greatly to the clock Allan deviation that averages down until 2000~s, at the level of 3~$\times$~10$^{-12}$. It is also interesting to note that, with SABR, a correlation was observed between the lab temperature and the routinely measured phase jump $\varphi_c$. These results confirm further that the SABR method efficiently mitigates light-shifts in microcells and improves the mid-term stability of miniaturized clocks, without any significant deterioration of the short-term stability.\\
For $\tau >$ 3000 s, the Allan deviation remains degraded, reaching the level of about 2.5 $\times$ 10$^{-11}$ at 10$^5$~s. However, several arguments tend to exonerate light-shift effects. From Fig.~\ref{fig:fig3}, we calculate that this stability level should be explained by fluctuations at 1 day of laser power and laser frequency, such that $\Delta P$~=~7.6 $\mu$W ($\Delta P/P =$~ 10.6~\%), and $\Delta f$ = 177 MHz ($\Delta f/f=$~5.2~$\times$~10$^{-7}$), respectively. These values are high in comparison with those reported in the literature \cite{Zhang:JOSAB:2016, Gruet:OLE:2013, Vicarini:SA:2018}. Variations $\Delta P_{\mu W}$ of the microwave power of only 0.06 dB at 1 day might be more suspected \cite{Zhang:JOSAB:2016, Vicarini:SA:2018, Yanagimachi:APL:2020} to justify the stability limitation. However, this assumption is counterbalanced by two observations. First, Allan deviation results obtained at 1 day are comparable for SABR and Ramsey-CPT cases whereas light-shift coefficients in the Ramsey case are at least one order of magnitude higher. Second, in both Ramsey and SABR cases, we observed that the clock frequency constantly drifted with a negative slope. This similar frequency drift sign, observed in both regimes, is contradictory with the fact that light-shift coefficients reported in Fig. ~\ref{fig:fig3}(b), obtained in Ramsey and SABR cases, exhibit opposite signs. We have also checked that the temperature dependence of the buffer gas collisional shift \cite{Miletic:EL:2010, Kozlova:PRA:2011}, the Zeeman shift, barometric effects \cite{Moreno:2018} or alkali condensation on the cell windows \cite{Karlen:EFTF:2018} could not explain the measured stability level at 1~day. To date, the clock stability limitation at 1 day on this setup is suspected to come from a possible unstable cell inner atmosphere. This could be related to Ne buffer gas permeation through the glass windows \cite{Abdullah:APL:2015}, materials degassing, or dispenser pollution \cite{HOPGs}. \\

In conclusion, we have explored the implementation of Auto-Balanced Ramsey (ABR) spectroscopy, for light-shift mitigation, in a microcell-based atomic clock. Despite the use of patterns with optical pulses and dark time durations of only a few hundreds of microseconds, this technique, provided that symmetry of the pattern is well applied, reduces the dependence of the clock frequency to laser field parameters variations by more than two orders of magnitude, with respect to the standard CW-regime approach, commonly used in commercial CSACs. We also demontrated that SABR contributes to improve the clock Allan deviation of a microcell CPT clock between 10$^2$ and 10$^5$ s, where light-shifts are predominant.\\ 

This work was partly funded by the Délégation Générale de
l’Armement (DGA), Centre National des Etudes Spatiales (CNES), Agence Innovation Défense (AID), Conseil R\'egional Bourgogne Franche-Comté with the HACES project (grant 2018-04768), and in part by Agence Nationale de la Recherche (ANR) in the frame of the LabeX FIRST-TF (Grant ANR 10-LABX-48-01), EquipX Oscillator-IMP (Grant ANR 11-EQPX-0033), ASTRID PULSACION (Grant ANR-19-ASTR-0013-01) projects and EIPHI Graduate school (Grant ANR-17-EURE-0002). This work was partly supported by the french RENATECH network and FEMTO-ST technological facility (MIMENTO). The authors thank deeply J. P. McGilligan (Strathclyde University) for careful reading of the manuscript.

\section*{Data availability statement}
The data supporting the findings of this study are available
from the corresponding author upon reasonable request.\\


\end{document}